\documentclass[a4paper,11pt]{article}

\usepackage{amssymb}
\usepackage{amsmath, bbm}
\usepackage[figuresright]{rotating}
\usepackage{capt-of}
\usepackage{multirow}

\usepackage{float} 

\usepackage{jheppub}
\allowdisplaybreaks

\newcommand{\ccc}{\cdot\cdot\cdot}
\newcommand{\beqa}{\begin{eqnarray}}
\newcommand{\eeqa}{\end{eqnarray}}

\newlength{\feynwidth} 

\title{Antinucleon-nucleon interaction in chiral effective field theory}
\author[a]{Xian-Wei Kang,}
\author[a]{Johann Haidenbauer,}
\author[b,a]{Ulf-G. Mei{\ss}ner}
\affiliation[a]{Institute for Advanced Simulation, Institut f{\"u}r Kernphysik and
J\"ulich Center for Hadron Physics, Forschungszentrum J{\"u}lich, D-52425 J{\"u}lich, Germany}
\affiliation[b]{Helmholtz Institut f\"ur Strahlen- und Kernphysik and Bethe Center
 for Theoretical Physics,\\ Universit\"at Bonn, D-53115 Bonn, Germany}

\emailAdd{x.kang@fz-juelich.de}
\emailAdd{j.haidenbauer@fz-juelich.de}
\emailAdd{meissner@hiskp.uni-bonn.de}

\bigskip

\abstract{ Results of an exploratory study of the antinucleon-nucleon interaction within
chiral effective field theory are reported. The antinucleon-nucleon potential 
is derived up to next-to-next-to-leading order, based on a modified Weinberg power
counting, in close analogy to pertinent studies of the nucleon-nucleon interaction.
The low-energy constants associated with the arising contact interactions 
are fixed by a fit to phase shifts and inelasticities provided by a recently
published phase-shift analysis of antiproton-proton scattering data.
The overall quality of the achieved description of the 
antinucleon-nucleon amplitudes is comparable to the one found in case of the
nucleon-nucleon interaction at the same order.
For most $S$-waves and several $P$-waves 
good agreement with the antinucleon-nucleon phase shifts and inelasticities 
is obtained up to laboratory energies of around 200 MeV. }


\begin{document}

\maketitle
\flushbottom

\section{Introduction}
\label{sec:1}

The antinucleon-nucleon ($\bar NN$) interaction has been studied 
quite extensively in the past 
\cite{Dover80,Dover82,Cote,Timmers,Hippchen,Mulla,Mull,Entem06,Bennich},
not least because of the wealth of data collected at the LEAR facility
at CERN, cf. the reviews \cite{Rev1,Rev2,Rev3}.
The majority of those investigations has been performed in the traditional
meson-exchange framework where the $G$-parity transformation is
exploited to connect the elastic part of the $\bar NN$ interaction
with the dynamics in the nucleon-nucleon ($NN$) system. Annihilation 
processes are described either by a simple optical potential (which is 
often assumed to be spin- as well as energy-independent) 
\cite{Dover80,Dover82,Hippchen,Mull} or in terms of a coupling
to a small number of effective two-body annihilation channels 
\cite{Cote,Timmers,Bennich}. 

In the last two decades chiral effective field theory (EFT) has 
become a standard tool in the studies of the $NN$ interaction at 
low energies.
This developement was initiated by two seminal papers by
Weinberg \cite{Wei90,Wei91} in which he proposed that
EFT and the power-counting rules associated with it should 
be applied to the $NN$ potential rather than to the reaction amplitude.
The reaction amplitude is then obtained from solving a regularized
Lippmann-Schwinger equation for the derived interaction potential.
His suggestion is based on the observation that diagrams with
purely nucleonic intermediate states are strongly enhanced and,
therefore, not amenable to a perturbative treatment. However,
they can be taken into account and they are actually summed up 
to infinite order when solving the Lippmann-Schwinger equation. 
The chiral $NN$ potential contains pion exchanges and a series of
contact interactions with an increasing number of derivatives. 
The latter represent the short-range part of the $NN$ force and are 
parametrized by low-energy constants (LECs), that need to be fixed by 
a fit to data. For reviews we refer the reader
to the recent Refs.~\cite{Epelbaum:2008ga,Machleidt:2011zz}.
Presently the most refined calculations extend up to 
next-to-next-to-next-to-leading order (N$^3$LO) \cite{Entem:2003ft,Epe05}
and they yield a rather accurate description of the $NN$ phase 
shifts up to laboratory energies of 250-300 MeV. 

Naturally, the success of chiral EFT in the $NN$ sector provides
a strong motivation to apply the same approach also to the $\bar NN$ 
interaction. First and most important for the practical
implementation, recently an update of the Nijmegen 
partial-wave analysis (PWA) of antiproton-proton ($\bar pp$) scattering data 
\cite{Timmermans} has been published. For the new PWA \cite{Zhou2012} 
the resulting phase shifts and inelasticities are explicitly given and can 
be readily used for applying the chiral EFT approach to the $\bar NN$ 
interaction in the very same way as it has been done for the $NN$ system. 

A further incentive for exploring the feasibility of investigating the
$\bar NN$ system within chiral EFT comes from the expected increase 
in interest in the $\bar NN$ interaction in the future due to the Facility 
for Antiproton and Ion Research (FAIR) in Darmstadt whose construction 
is finally on its way. Among the various project planned at this 
site is the PANDA experiment \cite{PANDA} which aims to study the 
interactions between antiprotons
and fixed target protons and nuclei in the momentum range of
1.5-15~GeV/c using the high energy storage ring HESR.

Finally, chiral EFT could be a very powerful tool to analyze data
from recent measurements of the $\bar pp$ invariant 
mass in the decays of $J/\psi$, $B$ mesons, etc., and of the
reaction $e^+e^- \to \bar pp$. 
In several of those reactions a near-threshold enhancement in 
the mass spectrum was found \cite{Bai,Aubert,Aubert1,BES12} and 
this enhancement could allow one to extract information on the
$\bar pp$ interaction at very low energies 
\cite{BuggFSI,Zou,Sibirtsev05,Loiseau,JH06,JH06a,Entem,Dedonder,JH12}.

In the present paper we report on results of an exploratory study of the 
antinucleon-nucleon interaction within chiral EFT. 
In our application of chiral EFT to the $\bar N N$ interaction we 
follow exactly the approach used by Epelbaum et 
al.~\cite{Epe05,Epe04,Epe04a} in the $NN$ case. 
It is consistent with the scheme originally proposed by Weinberg
except that one aims for an energy-independent representation of 
the chiral potential \cite{Epe98}. 
For the time being we restrict ourselves to an evaluation 
of the potential up to next--to--next--to--leading order (NNLO).
At leading order (LO) the potential is given by one--pion exchange 
(OPE) and two contact terms without derivatives.
At next--to--leading order (NLO) contributions from the 
leading two--pion exchange (TPE) diagrams as well as seven more 
contact operators arise. Finally, at NNLO one gets contributions
from the subleading TPE with one insertion of dimension two 
pion--nucleon vertices.
Once the potential is established it has to be inserted into a
regularized scattering equation in order to obtain the reaction
amplitude. For the regularization we follow again closely the 
procedure adopted by Epelbaum et 
al.~\cite{Epe05,Epe04a} and others \cite{Entem:2003ft}, in their 
study of the $NN$ interaction and introduce a momentum-dependent 
exponential regulator function. 

For investigations of the $\bar NN$ interaction within EFT
based on other schemes see Refs.~\cite{Chen2010,Chen2011}, where the
Kaplan-Savage-Wise resummation scheme \cite{Kaplan} is employed.
These authors considered the $\bar NN$ interaction up to NLO.
There have been also attempts to compute specific $\bar pp$
annihilation channels in chiral EFT \cite{Tarasov}. 

The present paper is structured as follows:
The effective $\bar NN$ potential up to NNLO is described in Section 2.
We start with a brief review of the underlying power counting
and then provide explicit expressions for the contributions from
pion exchange and for the contact terms. 
We also discuss how we treat the annihilation processes. 
Finally, we introduce the Lippmann-Schwinger equation that we 
solve and the parameterization of the S-matrix that we use. 
In Section 3 we indicate our fitting procedure and then we
present the results achieved at NLO and at NNLO. Phase shifts
and inelasticites for $S$-, $P$-, and $D$- waves, obtained
from our EFT interaction, are displayed and compared with 
those of the $\bar NN$ phase-shift analysis.
Furthermore, predictions for $S$-wave scattering lengths are
given. 
A summary of our work and an outlook on future investigations
is given in Section~4. 

\section{Chiral potential at next-to-next-to-leading order}
\label{sec:2}

\begin{figure}[h]
 \centering
 \includegraphics[width=\feynwidth]{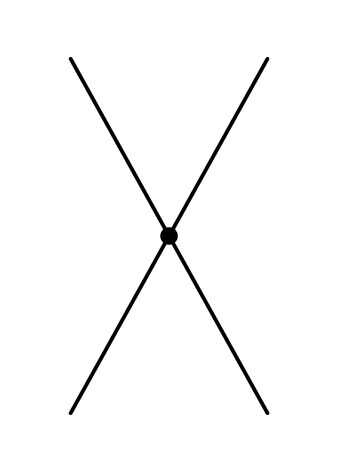}\includegraphics[width=\feynwidth]{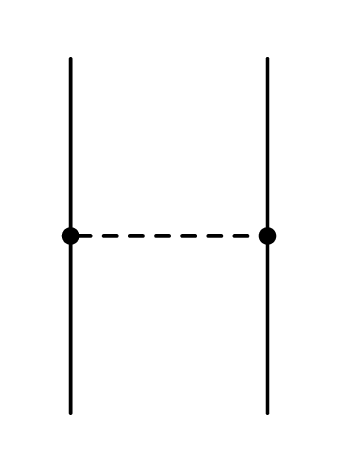}

 \includegraphics[width=\feynwidth]{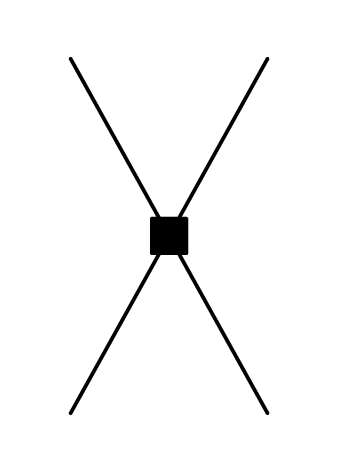}
 \includegraphics[width=\feynwidth]{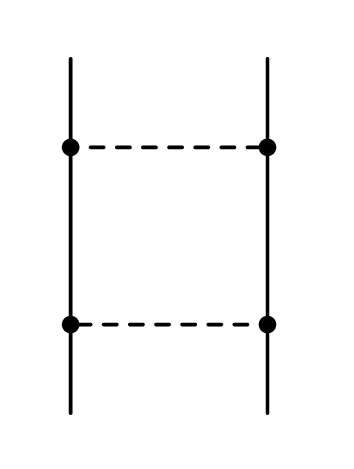}
 \includegraphics[width=\feynwidth]{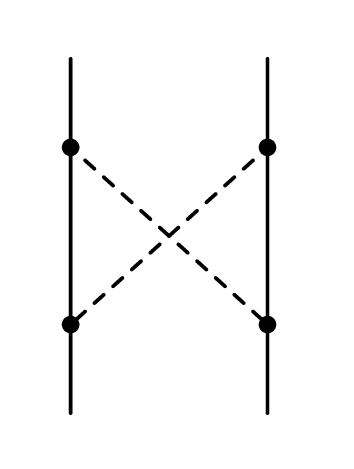}
 \includegraphics[width=\feynwidth]{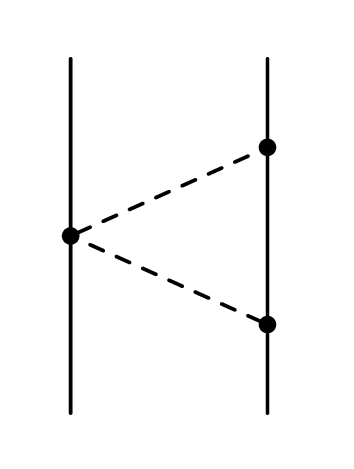}
 \includegraphics[width=\feynwidth]{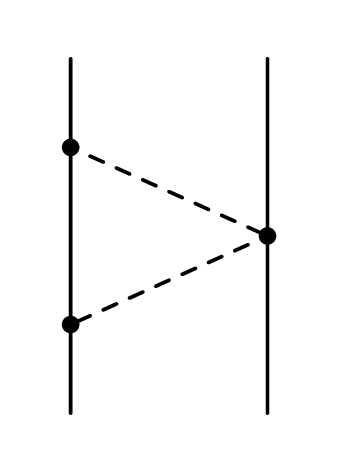}
 \includegraphics[width=\feynwidth]{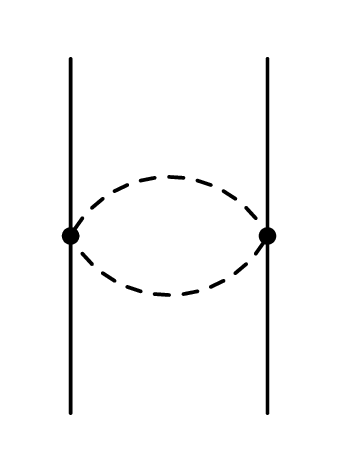}

\includegraphics[width=\feynwidth]{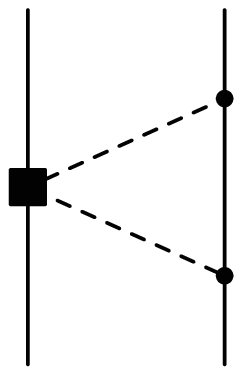}
\includegraphics[width=\feynwidth]{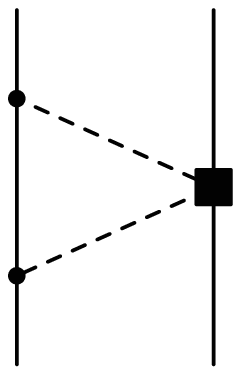}
\includegraphics[width=\feynwidth]{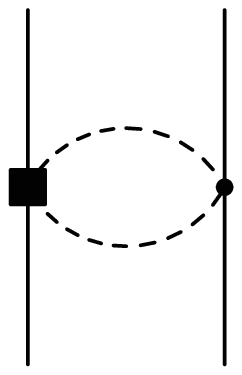}
\includegraphics[width=\feynwidth]{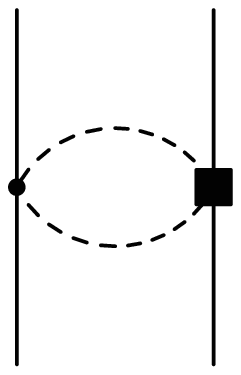}

\caption{Relevant diagrams up-to-and-including NNLO. 
Solid and dashed lines denote the antinucleon/nucleon and the pion, respectively. 
The square symbolizes a contact vertex with two derivatives or a subleading $\pi N$ vertex. 
The contributions at LO, NLO, and NNLO are displayed from top to bottom.
} \label{fig:feynman}
\end{figure}

The contributions to the $NN$ interaction up to NNLO are described in detail
in Refs.~\cite{Epe05,Epe04,Epe04a}. The underlying power counting is given 
by (considering only connected diagrams) 
\begin{equation}
\nu = 2L +\sum_i\Delta_i, \quad \Delta_i=d_i+\frac{n_i}{2}-2
\end{equation}
where $L$ is the number of loops in the diagram, $d_i$ is the number of derivatives 
or pion mass insertions, and $n_i$ the number of internal nucleon fields at the 
vertex $i$ under consideration. 
The LO potential corresponds to $\nu=0$ and consists of two four-nucleon 
contact terms without derivatives and of one-pion exchange. 
There are no contributions at order $\nu=1$ due to requirements from parity 
conservation and time-reversal invariance. 
At NLO ($\nu=2$) seven new contact terms (with two derivatives)
arise, together with loop contributions from (irreducible) two-pion exchange.
Finally, at NNLO ($\nu=3$) there are additional contributions from two-pion exchange
resulting from one insertion of dimension two pion-nucleon vertices, see
e.g. Ref.~\cite{Bernard:1995dp}.
The corresponding diagrams are summarized in Fig.~\ref{fig:feynman}.
  
The structure of the $\bar NN$ interaction is practically identical and,
therefore, the potential given in Refs.~\cite{Epe05,Epe04a} can be adapted 
straightforwardly for the $\bar NN$ case. For the ease
of the reader and also for defining our potential uniquely we provide
the explicit expressions below.

\subsection{Pion exchange}
\label{sec:2MEX}

In line with \cite{Epe05} we adopt the following expression for the one-pion exchange potential 
\begin{equation}
\label{opep_full}
V_{1\pi} (q) =  \biggl(\frac{g_A}{2F_\pi}\biggr)^2 \, 
\left( 1 - \frac{p^2 + p'^2}{2 m^2} \right) 
\mbox{\boldmath $\tau$}_1 \cdot \mbox{\boldmath $\tau$}_2 \, 
\frac{\mbox{\boldmath $\sigma$}_1 \cdot{\bf q}\,\mbox{\boldmath$\sigma$}_2\cdot{\bf q}}
{{\bf q}^2 + M_\pi^2} \ ,
\end{equation}
where ${\bf q}={\bf p}'-{\bf p}$ is the transferred momentum defined in terms of the 
final (${\bf p}'$) and initial (${\bf p}$) center-of-mass momenta of the baryons 
(nucleon or antinucleon). 
Obviously here relativistic $1/m^2$ corrections to the static one-pion exchange 
potential have been taken into account. As in the work \cite{Epe05} 
we take the larger value $g_A = 1.29$ instead of $g_A = 1.26$ in order to 
account for the Goldberger--Treiman discrepancy.  This value, together with
the used $F_\pi = 92.4$ MeV, implies the pion-nucleon coupling 
constant $g_{NN\pi} =13.1$ which is consistent with the empirical 
value obtained from $\pi N$ and $NN$ data \cite{deSwart,Bugg} and also
with modern determinations utilizing the GMO sum rule \cite{Baru:2011bw}. 
For the nucleon (antinucleon) and pion mass we use the isospin-averaged
values $m=938.918$~MeV and $M_\pi=138.039$~MeV, respectively. 
Note that the contribution of
one-pion exchange to the $\bar NN$ interaction is of opposite sign as that in
the $NN$ case. This sign difference arises from transforming the $NN\pi$ vertex
to the $\bar N \bar N\pi$ vertex via charge conjugation and a rotation
in the isospin space and is commonly referred to as $G$-parity transformation. 

The two-pion exchange potential calculated using spectral function regularization 
\cite{Epe05} is given at NLO by 
\begin{equation}
\label{tpep2}
V_{2\pi}^{(2)} (q) = 
\mbox{\boldmath $\tau$}_1 \cdot \mbox{\boldmath $\tau$}_2 \, V_C^{(2)} (q) + 
{\mbox{\boldmath $\sigma$}_1 \cdot{\bf q}\,\mbox{\boldmath$\sigma$}_2\cdot{\bf q}}
\, V_T^{(2)} (q) + 
{\mbox{\boldmath $\sigma$}_1 \cdot\,\mbox{\boldmath$\sigma$}_2}\, V_S^{(2)} (q) ~,
\end{equation}
where 
\beqa
\label{2PE_nlo}
V_C^{(2)} (q) &=& - \frac{1}{384 \pi^2 F_\pi^4}\,
L^{\tilde \Lambda} (q) \, \biggl\{4M_\pi^2 (5g_A^4 - 4g_A^2 -1)  + q^2(23g_A^4 - 10g_A^2 -1)
+ \frac{48 g_A^4 M_\pi^4}{4 M_\pi^2 + q^2} \biggr\} ~,\cr \nonumber
V_T^{(2)} (q) &=& -\frac{1}{q^2} V_S^{(2)} (q)  
= - \frac{3 g_A^4}{64 \pi^2 F_\pi^4} \,L^{\tilde \Lambda} (q)~,
\eeqa
and at NNLO by  
\begin{equation}
\label{tpep3}
V_{2\pi}^{(3)} (q) = 
V_C^{(3)} (q) + 
\mbox{\boldmath $\tau$}_1 \cdot \mbox{\boldmath $\tau$}_2 \, 
{\mbox{\boldmath $\sigma$}_1 \cdot{\bf q}\,\mbox{\boldmath$\sigma$}_2\cdot{\bf q}}
\, V_T^{(3)} (q) + 
\mbox{\boldmath $\tau$}_1 \cdot \mbox{\boldmath $\tau$}_2 \, 
{\mbox{\boldmath $\sigma$}_1 \cdot\,\mbox{\boldmath$\sigma$}_2}\, V_S^{(3)} (q) ~,
\end{equation}
with 
\beqa
\label{2PE_nnlo}
V_C^{(3)} (q) &=& -\frac{3g_A^2}{16\pi F_\pi^4}  \biggl\{2M_\pi^2(2c_1 -c_3) -c_3 q^2 \biggr\} 
(2M_\pi^2+q^2) A^{\tilde \Lambda} (q) ~, \cr \nonumber
V_T^{(3)} (q) &=& -\frac{1}{q^2} V_S^{(3)} (q) = - \frac{g_A^2}{32\pi F_\pi^4} \,  c_4 (4M_\pi^2 + q^2) 
A^{\tilde \Lambda}(q)\, .
\eeqa
The NLO and NNLO loop functions $L^{\tilde \Lambda} (q)$ and $A^{\tilde \Lambda} (q)$
are given by 
\begin{equation}
\label{def_LA}
L^{\tilde \Lambda} (q) = \theta (\tilde \Lambda - 2 M_\pi ) \, \frac{\omega}{2 q} \, 
\ln \frac{\tilde \Lambda^2 \omega^2 + q^2 s^2 + 2 \tilde \Lambda q 
\omega s}{4 M_\pi^2 ( \tilde \Lambda^2 + q^2)}~, \quad 
\omega = \sqrt{ q^2 + 4 M_\pi^2}~,  \quad 
s = \sqrt{\tilde \Lambda^2 - 4 M_\pi^2}\, ,
\end{equation}
and
\begin{equation}
A^{\tilde \Lambda} (q) = \theta (\tilde \Lambda - 2 M_\pi ) \, \frac{1}{2 q} \, 
\arctan \frac{q ( \tilde \Lambda - 2 M_\pi )}{q^2 + 2 \tilde \Lambda M_\pi}\,.
\end{equation}

For the LECs $c_{1}$ and $c_{4}$ we adopt the central values
from the $Q^3$--analysis of the $\pi N$ system \cite{Paul}:
$c_1=-0.81$ GeV$^{-1}$, $c_4=3.40$ GeV$^{-1}$. 
For the constant $c_3$ the value $c_3=-3.40$~GeV$^{-1}$ is used, which is on 
the lower side but still consistent with the results from Ref.~\cite{Paul}.
Note that slightly different values are employed in the $\bar NN$ partial-wave 
analysis \cite{Zhou2012}, namely 
$c_1=-0.76$ GeV$^{-1}$, $c_3=-5.8$ GeV$^{-1}$ and $c_4=4.0$ GeV$^{-1}$.
These values are also consistent with the recent determination in
\cite{Krebs:2012yv}.

\subsection{Contact terms}
\label{sec:2CT}

The spin-dependence of the potentials due to the leading order contact terms is given by \cite{Epe00}
\begin{eqnarray}
V^{(0)}_{\bar NN} &=& C_{S} + C_{T}\,
\mbox{\boldmath $\sigma$}_1\cdot\mbox{\boldmath $\sigma$}_2\,,
\end{eqnarray}
where the parameters $C_{S}$ and $C_{T}$ are low-energy 
constants (LECs) which need to be determined in a fit to data. 
At NLO, the spin- and momentum-dependence of the contact terms reads
\begin{eqnarray}
V^{(2)}_{\bar NN} &=& C_1 {\bf q}^{\,2}+ C_2 {\bf k}^{\,2} + (C_3 {\bf q}^{\,2}+ C_4 {\bf k}^{\,2})
\,\mbox{\boldmath $\sigma$}_1\cdot\mbox{\boldmath $\sigma$}_2 
+ \frac{i}{2} C_5 (\mbox{\boldmath $\sigma$}_1+\mbox{\boldmath $\sigma$}_2)\cdot ({\bf q} \times {\bf k}) \nonumber \\
&+& C_6 ({\bf q} \cdot \mbox{\boldmath $\sigma$}_1) ({\bf q} \cdot \mbox{\boldmath $\sigma$}_2)
+ C_7 ({\bf k} \cdot \mbox{\boldmath $\sigma$}_1) ({\bf k} \cdot \mbox{\boldmath $\sigma$}_2) \ ,
\end{eqnarray}
where $C_i$ ($i=1,\dots,7$) are additional LECs.
The average momentum ${\bf k}$ is defined by ${\bf k}=({\bf p}'+{\bf p})/2$. 
When performing a partial-wave projection, these terms contribute to the two $S$--wave
($^1S_0$, $^3S_1$) potentials, the four $P$--wave
($^1P_1$, $^3P_0$, $^3P_1$, $^3P_2$) potentials, and the $^3S_1$-$^3D_1$ 
transition potential in the following way \cite{Epe05}:
\begin{eqnarray}
\label{VC0}
V(^1S_0) &=& {4\pi} \, (C_S-3C_T) + \pi \, ( 4C_1 + C_2 -12C_3
-3C_4 -4C_6 -C_7) ({p}^2+{p}'^2)\, \nonumber \\
&=& \tilde{C}_{^1S_0} + {C}_{^1S_0} ({p}^2+{p}'^2)\,, \label{C1S0}\\
V(^3S_1) &=& {4\pi} \, (C_S+C_T) + \frac{\pi}{3} \, ( 12C_1 + 3C_2 +12C_3
+3C_4 +4C_6 +C_7) ({p}^2+{p}'^2)\,\nonumber \\
&=& \tilde{C}_{^3S_1} + {C}_{^3S_1} ({p}^2+{p}'^2)~, \label{C3S1} \\
V(^1P_1) &=& \frac{2\pi}{3} \, ( -4C_1 + C_2 +12C_3
-3C_4 +4C_6 -C_7) \, {p}\, {p}'
= {C}_{^1P_1}\, {p}\, {p}'\,,\\
V(^3P_1) &=& \frac{2\pi}{3} \, ( -4C_1 + C_2 - 4C_3
+C_4 + 2C_5 -8C_6 +2C_7) \, {p}\, {p}'
= {C}_{^3P_1}\, {p}\, {p}'\,, \\ 
V(^3P_0) &=& \frac{2\pi}{3} \, ( -4C_1 + C_2 - 4C_3
+C_4 + 4C_5 +12C_6 - 3C_7) \, {p}\, {p}'
= {C}_{^3P_0}\, {p}\, {p}'\,,\\
V(^3P_2) &=& \frac{2\pi}{3} \, ( -4C_1 + C_2 - 4C_3
+C_4 - 2C_5 ) \, {p}\, {p}'
= {C}_{^3P_2}\,  {p}\, {p}'\,,\\
V(^3D_1 -\, ^3S_1) &=& \frac{2\sqrt{2}\pi}{3} \, ( 4C_6 + C_7)\,
{p'}^2 = {C}_{^3S_1 -\, ^3D_1}\, {p'}^2 \equiv {C}_{\epsilon_1}\, {p'}^2 \,,\\
V(^3S_1 -\, ^3D_1) &=& \frac{2\sqrt{2}\pi}{3} \, ( 4C_6 + C_7)\,
{p}^2 = {C}_{^3S_1 -\, ^3D_1}\, {p}^2\equiv {C}_{\epsilon_1}\, {p}^2\,, 
\label{VC}
\end{eqnarray}
with $p = |{\bf p}\,|$ and ${p}' = |{\bf p}\,'|$.
There are no additional contact terms at NNLO. 

Note that the Pauli principle is absent in case of the $\bar NN$ interaction.
Accordingly, each partial wave that is allowed by angular momentum conservation 
occurs in the isospin $I=0$ and in the $I=1$ channel.
Therefore, there are now twice as many contact terms as in $NN$.

The main new feature in the $\bar NN$ interaction is the presence of 
annihilation processes. The $\bar NN$ system annihilates into a multitude
of $n \, \pi$ channels, where the decay to 4 to 6 pions is dominant in the
low-energy region of $\bar NN$ scattering \cite{Rev1}. 
The threshold energy of those 
channels is in the order of 700~MeV while the $\bar NN$ threshold is at 1878~MeV.
Therefore, one does not expect that annihilation introduces a new scale into
the problem. Accordingly, there should be no need to modify the power 
counting when going from $NN$ to $\bar NN$ because the momenta associated with 
the annihilation channels should be, in average, much larger than those in 
the $\bar NN$ system itself. 
This conjecture is supported by the fact that phenomenological models of the
$\bar NN$ interaction can describe the bulk properties of annihilation very
well by simple energy-independent optical potentials of Woods-Saxon or
Gaussian type \cite{Dover80,Dover82,Hippchen,Mull}. 
The ranges associated with those interactions are of the
order of 1~fm or less. The above considerations suggest that annihilation 
processes are primarily tied to short-distance physics and, therefore, can be 
and should be simply incorporated into the contact terms which anyway are meant
to parameterize effectively the short-range part of (elastic) $NN$ and/or 
$\bar NN$ scattering. 

Nonetheless we want to emphasize that the above arguments are of pragmatical 
nature and not fundamental ones. There are definitely annihilation channels 
that open near the $\bar NN$ threshold. Specifically, there are indications 
that a sizeable part of the annihilation into multipion channels proceeds 
via two-meson doorway modes like $\bar NN \to \rho\rho \to 4\pi$ or 
$\bar NN \to f_2(1270)\omega \to 5\pi$, and some of those have
nominal thresholds close to that of $\bar NN$ scattering.  
On the other hand, according to empirical information the actual branching 
ratios into individual two-body channels are typically of the order 
of 1\% \cite{Mull} only and, therefore, 
they do not have any noticeable impact on the description of the bulk properties of 
$\bar NN$ annihilation. In fact, all the two-body annihilation channels together 
-- as far as they have been measured -- yield only about 30\% of the total
annihilation cross section at the $\bar NN$ threshold which is a strong
evidence for the dominance of annihilation into 3 or more (uncorrelated) pions. 

The study of $\bar NN$ scattering in EFT in Refs.~\cite{Chen2010,Chen2011}
followed the above arguments and took into account annihilation by simply
using complex LECs in Eqs.~(\ref{VC0})-(\ref{VC}). 
However, this prescription has an unpleasant drawback -- it does not allow one
to impose sensible unitarity requirements on the resulting scattering amplitude. 
With unitarity requirements we mean a condition that guarantees 
that for each partial wave its contribution to the total cross section
is larger than its contribution to the integrated elastic cross section.
In case of strict two-body unitary like for $NN$ scattering below the
pion production threshold these two quantities are, of course, identical.

Since we want an approach that manifestly fulfils unitarity
constraints we treat annihilation in a different way. We start out from
the observation that unitarity requires the $\bar NN$ annihilation
potential to be of the form 
\begin{equation}
V_{ann} = \sum_{X=2\pi,3\pi,...} V_{\bar NN \to X} G_X^{} V_{X\to \bar NN} 
\label{ANN} 
\end{equation}
where $X$ is the sum over all open annihilation channels, and $G_X$ is the
propagator of the intermediate state $X$. 
Note that Eq.~(\ref{ANN}) is exact under the assumption that there is no 
interaction in and no transition between the various annihilation channels.
Performing an expansion of $V_{\bar NN \to X}$ up to NNLO analoguous to 
the $\bar NN$ interaction and evaluating formally the sum and integral in 
Eq.~(\ref{ANN}) yields a contribution from the unitarity cut that can be
written as 
\begin{equation}
V^{L=0}_{ann} = -i\, (\tilde C_{^1S_0}^a+C_{^1S_0}^ap^2)(\tilde C_{^1S_0}^a+C_{^1S_0}^ap'^2), \ \ \ \  
V^{L=1}_{ann} = -i\, (C_\alpha^a)^2 p p',  
\label{ANN1} 
\end{equation}
where $\alpha$ stands for the $^3P_0$, $^1P_1$, $^3P_1$, and $^3P_2$ partial waves. 
For the coupled $^3S_1-^3D_1$ partial wave we get 
\begin{eqnarray}
V^{S\to S}_{ann} &=& -i\, (\tilde C_{^3S_1}^a+C_{^3S_1}^ap^2)(\tilde C_{^3S_1}^a+C_{^3S_1}^ap'^2), \ \ \ \
V^{S\to D}_{ann} = -i\, (\tilde C_{^3S_1}^a+C_{^3S_1}^ap^2)\, C_{\epsilon_1}^a p'^2, \nonumber \\
V^{D\to S}_{ann} &=& -i\,  C_{\epsilon_1}^a p^2\, (\tilde C_{^3S_1}^a+C_{^3S_1}^ap'^2), \ \ \ \ \ \ \ 
V^{D\to D}_{ann} = -i\,  (C_{\epsilon_1}^a)^2 p^2p'^2 \ .  
\label{ANN2} 
\end{eqnarray}
In those expressions the parameters $\tilde C^a$ and $C^a$ are real. Thus, 
for each partial wave we 
essentially recover the structure of the potential that follows from the contact 
terms considered above, with the same number of free parameters. However, 
in Eqs.~(\ref{ANN1})--(\ref{ANN2})
the sign of $V_{ann}$ as required by unitarity is already explicitly fixed
and does not depend on the sign of the parameters $\tilde C^a$ and $C^a$ anymore.
Moreover, and most importantly, we see that a term proportional to $p^2p'^2$ 
arises in the $S$ waves at NLO and NNLO from unitarity constraints and it has 
to be included in order to make sure that unitarity is fulfilled at any energy. 

Note that, in principle, there is also a contribution from the principal-value 
part of the integral in Eq.~(\ref{ANN}). However, it is real and, therefore, 
its structure is already accounted for by the standard LECs in 
Eqs.~(\ref{VC0})--(\ref{VC}). 

Finally we would like to add that in practice the treatment of annihilation 
via Eqs.~(\ref{ANN1})--(\ref{ANN2})
corresponds to the introduction of an effective two-body annihilation channel
with a threshold significantly below the one of $\bar NN$ so that the 
center-of-mass momentum in the annihilation channel is already fairly 
large and its variation in the low-energy region of $\bar NN$ scattering
considered by us is negligible. 

\subsection{Scattering equation}

In the actual calculation a partial-wave projection of the interaction 
potentials is performed, as described in detail in Ref.~\cite{Epe05}. 
The reaction amplitudes are obtained from the solution of a relativistic 
Lippmann-Schwinger (LS) equation: 
\begin{eqnarray}
&&T_{L''L'}(p'',p';E_q)=V_{L''L'}(p'',p')+
\sum_{L}\int_0^\infty \frac{dpp^2}{(2\pi)^3} \, V_{L''L}(p'',p)
\frac{1}{2E_q-2E_p+i0^+}T_{LL'}(p,p';E_q).\nonumber\\
\label{LS} 
\end{eqnarray}
Here, $E_q=\sqrt{m^2+q^2}$, where $q$ is the on-shell momentum.  
Like in the $NN$ case we have either uncoupled spin-singlet and triplet
waves (where $L''=L'=L=J$) or coupled partial waves 
(where $L'',L',L=J-1,J+1$). 
We solve the LS equation in the isospin basis, i.e. for $I=0$ and $I=1$ 
separately, and we compare the resulting phase shifts with those 
in Ref.~\cite{Zhou2012} that are likewise given in the isospin basis. 
It should be said, however, that for a comparison directly with
data a more refined treatment is required. 
Then one should solve the LS equation in particle basis and consider 
the coupling between the $\bar pp$ and $\bar nn$ channels explicitly. 
In this case one can take into account the mass 
difference between $p$ ($\bar p$) and $n$ ($\bar n$) and, thereby, 
implement the fact that the physical thresholds of the $\bar pp$ and 
$\bar nn$ channels are separated by about 2.5~MeV, 
and also one can add the Coulomb interaction in the $\bar pp$ channel. 
The potential in the LS equation is cut off with a regulator function, 
\begin{equation}
f^\Lambda(p',p) = \exp\left[-\left(p'^6+p^6\right)/\Lambda^6\right], 
\label{cutoff}
\end{equation}
in order to remove high-energy components \cite{Epe05}.
The cutoff values are chosen in the range $\Lambda=450$ -- $600\,$MeV at NLO
and $\Lambda=450$ -- $650\,$MeV at NNLO, similar to what 
was used for chiral $NN$ potentials \cite{Epe05,Epe04a}.

The relation between the  $S$-- and on--the--energy shell $T$--matrix is given by
\begin{equation}
\label{Sdef}
S_{L L'} (q) = \delta_{L L'} - \frac{i}{8 \pi^2}
\, q \, E_q \,  T_{L L'}(q)~.
\end{equation}
The phase shifts in the uncoupled cases can be obtained from the
$S$--matrix via
\begin{equation}
S_{LL} \equiv S_{L} = e^{ 2 i \delta_{L}} \, .
\label{SM0}
\end{equation}
For the $S$--matrix in the coupled channels ($J>0$) we 
use the so--called Stapp parametrization~\cite{Stapp}
\beqa
\left( \begin{array}{cc} S_{J-1 \, J-1} &  S_{J-1 \, J+1} \\
S_{J+1 \, J-1} &  S_{J+1 \, J+1} \end{array} \right) 
=
\left( \begin{array}{cc} \cos{2 \epsilon_J}\, e^{2 i \delta_{J-1}} &
-i \sin{2 \epsilon_J}\, e^{i(\delta_{J-1} + \delta_{J+1})} \\
-i \sin{2 \epsilon_J}\, e^{i(\delta_{J-1} + \delta_{J+1})} &
\cos{2 \epsilon_J}\, e^{2 i \delta_{J+1}} \end{array} \right)~.
\label{SM1}
\eeqa

In case of elastic scattering the phase parameters in Eqs.~(\ref{SM0}) 
and (\ref{SM1}) are real quantities while in the presence of inelasticites 
they become complex. Because of that, in the past several generalizations 
of these formulae have been proposed that still allow one to write the
$S$-matrix in terms of real parameters \cite{Arndt,Zhou2012}.
We follow here Ref.~\cite{Bystricky} and calculate and present simply
the real and imaginary parts of the phase shifts and the mixing
parameters obtained via the above parameterization. Note that with this
choice the real part of the phase shifts is identical to the phase shifts 
one obtains from another popular parameterization where the imaginary 
part is written in terms of an inelasticity parameter $\eta$, e.g. 
for uncoupled partial waves
\begin{equation}
S_{L} = \eta e^{ 2 i \delta_{L} } \; .
\label{SM2}
\end{equation}
Indeed, for this case ${\rm Im}\, \delta_{L} = -(\log \eta ) / 2$
which implies that ${\rm Im}\, \delta_{L} \geq 0$ 
since $\eta \leq 1$ because of unitarity. 
Since our calculation implements unitarity, the optical theorem
\begin{equation}
{\rm Im} \, a_{LL}(q) \geq q \, \sum_{L'} |a_{LL'}(q)|^2 \ , 
\label{OT} 
\end{equation}
is fulfilled for each partial wave, where
$ a_{LL'}(q) = (S_{LL'} - \delta_{LL'}) /2iq = 
- 1/(4\pi)^2 \, E_q \, T_{L L'}(q)$.

For the fitting procedure and for the comparison of our results
with those by Zhou and Timmermans we reconstructed the $S$-matrix
based on the phase shifts listed in Tables VIII-X in Ref.~\cite{Zhou2012}
and on the formulae presented in Sect. VII of that paper and then
converted them to our convention specified in Eqs.~(\ref{SM0}) and
(\ref{SM1}).

\section{Results}
In the fitting procedure we follow very closely the strategy of
Epelbaum et al. in their study of the $NN$ interaction \cite{Epe05,Epe04a}.
In particular, we consider the same ranges for the
cutoffs, namely for the cutoff in the LS equation values of
$\Lambda$ = 450--600~MeV at NLO and $\Lambda$ = 450--650~MeV at NNLO 
while for the spectral function regularization variations we consider 
values in the range $\tilde\Lambda$ = 500--700~MeV.
For any combination of the cutoffs $\Lambda$ and $\tilde \Lambda$,
the LECs $C_{S, T}$ and $C_{1 \ldots 7}$
are fixed from a fit to the $\bar NN$ $S$- and $P$-waves and the mixing parameter
$\epsilon_1$ of Ref.~\cite{Zhou2012} for laboratory energies below 
125~MeV ($p_{lab}\leq 500$~MeV/c). 
The numerical values of the LECs are compiled in Tables~\ref{tab:LEC1}
(NLO) and \ref{tab:LEC2} (NNLO) for a selected combination of the cutoffs. 
The values for $\tilde C_{^1S_0}$ in the isospin $I=1$ case found in the fitting
procedure turned out to be very small and, therefore, we set them to zero.

\begin{figure}[h]
\begin{center}
\includegraphics[height=85mm]{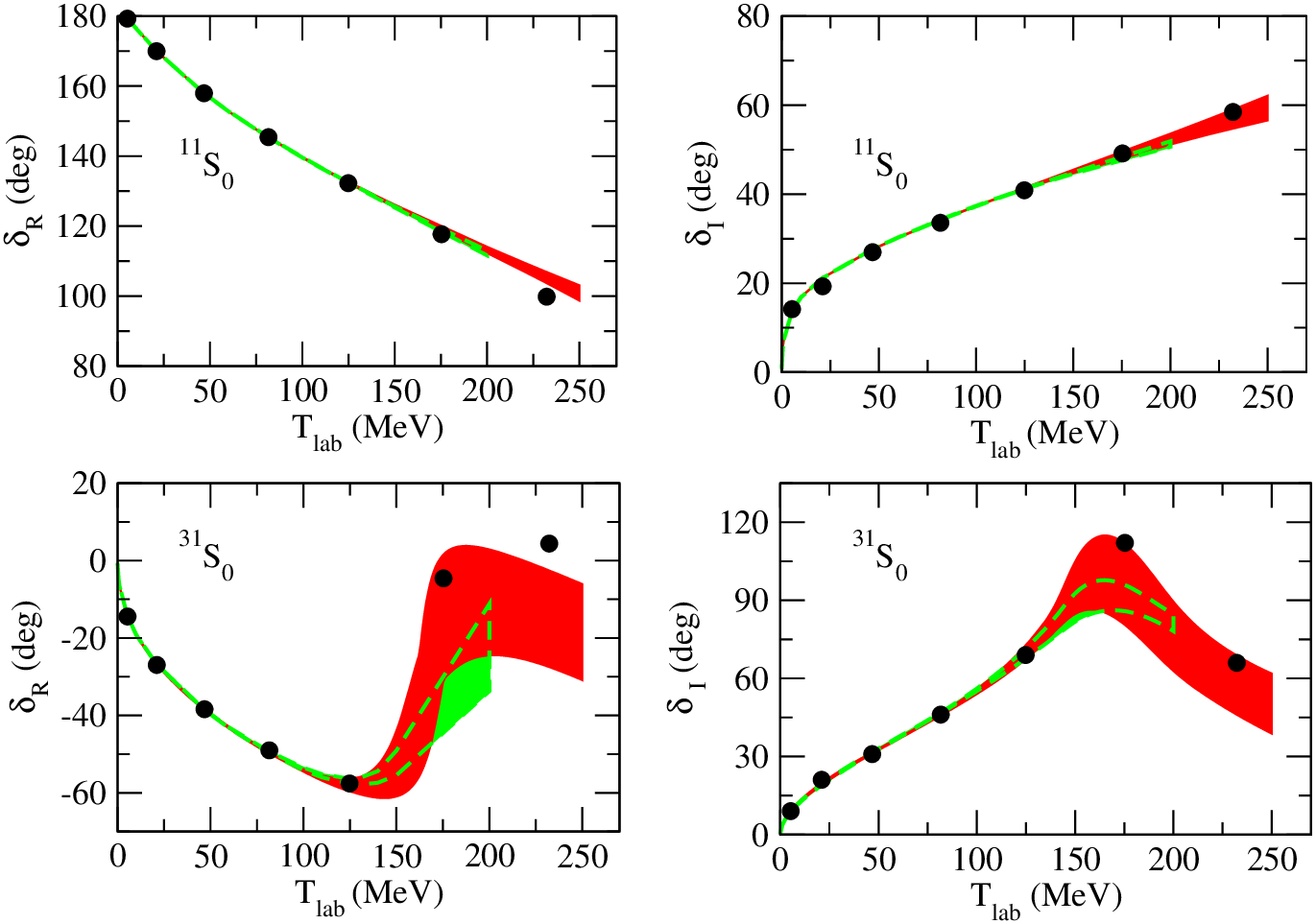}
\vskip 1.5cm
\includegraphics[height=85mm]{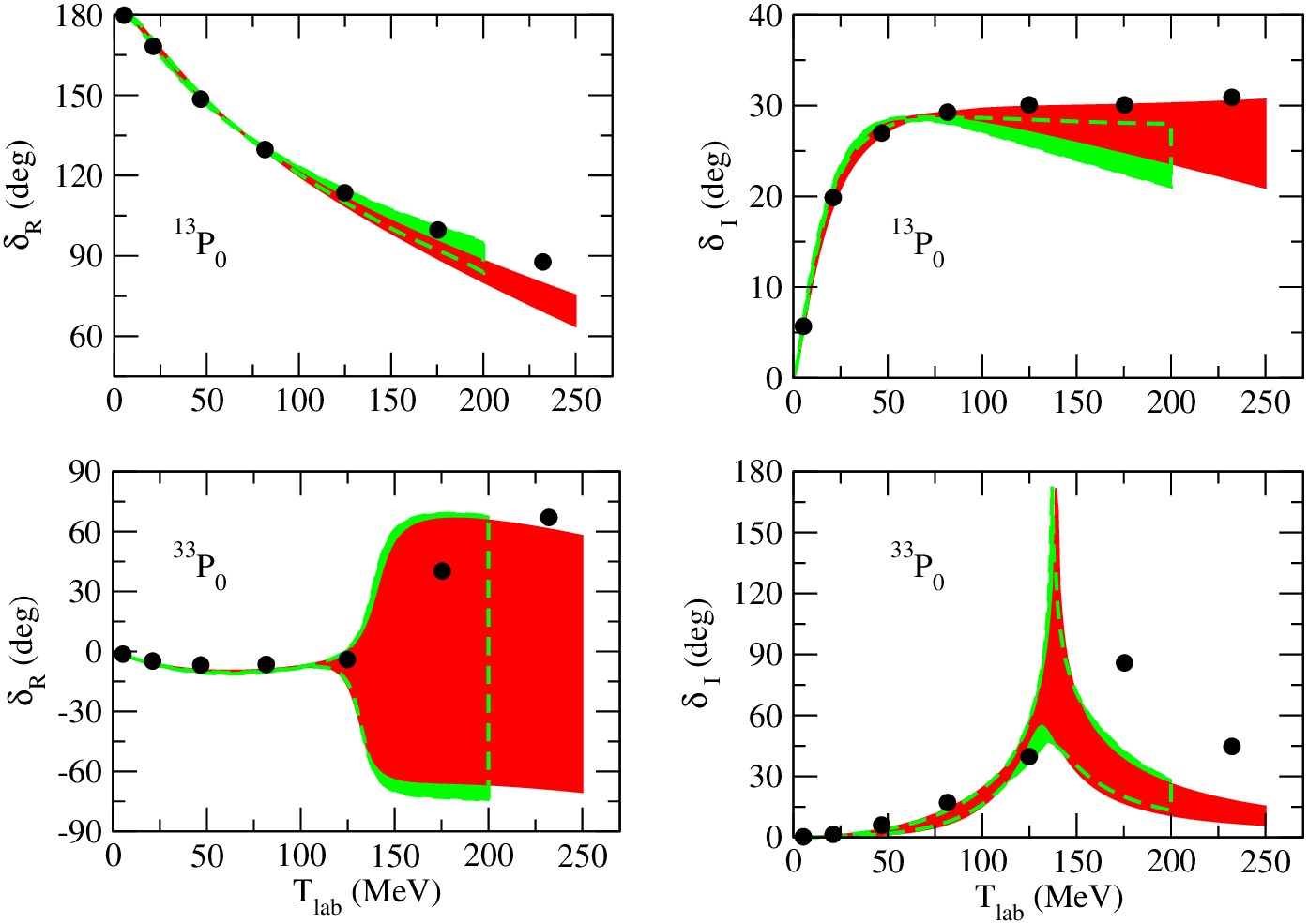}

\caption{Real and imaginary parts of the phase shift in the $^1S_0$ and $^3P_0$ partial waves.
The red/dark band shows the chiral EFT results up to NNLO for variations of the cutoff
in the range $\Lambda =$ 450--650~MeV in the Lippmann-Schwinger equation,
while the green/light band are results to NLO for $\Lambda =$ 450--600~MeV. The cutoff in the 
pion loops is varied independently in the range $\tilde \Lambda =$ 500--700~MeV. 
The solid circles represent the solution of the PWA of Ref.~\cite{Zhou2012}. 
}
\label{fig:1S0}
\end{center}
\end{figure}

\begin{figure}[h]
\begin{center}
\includegraphics[height=85mm]{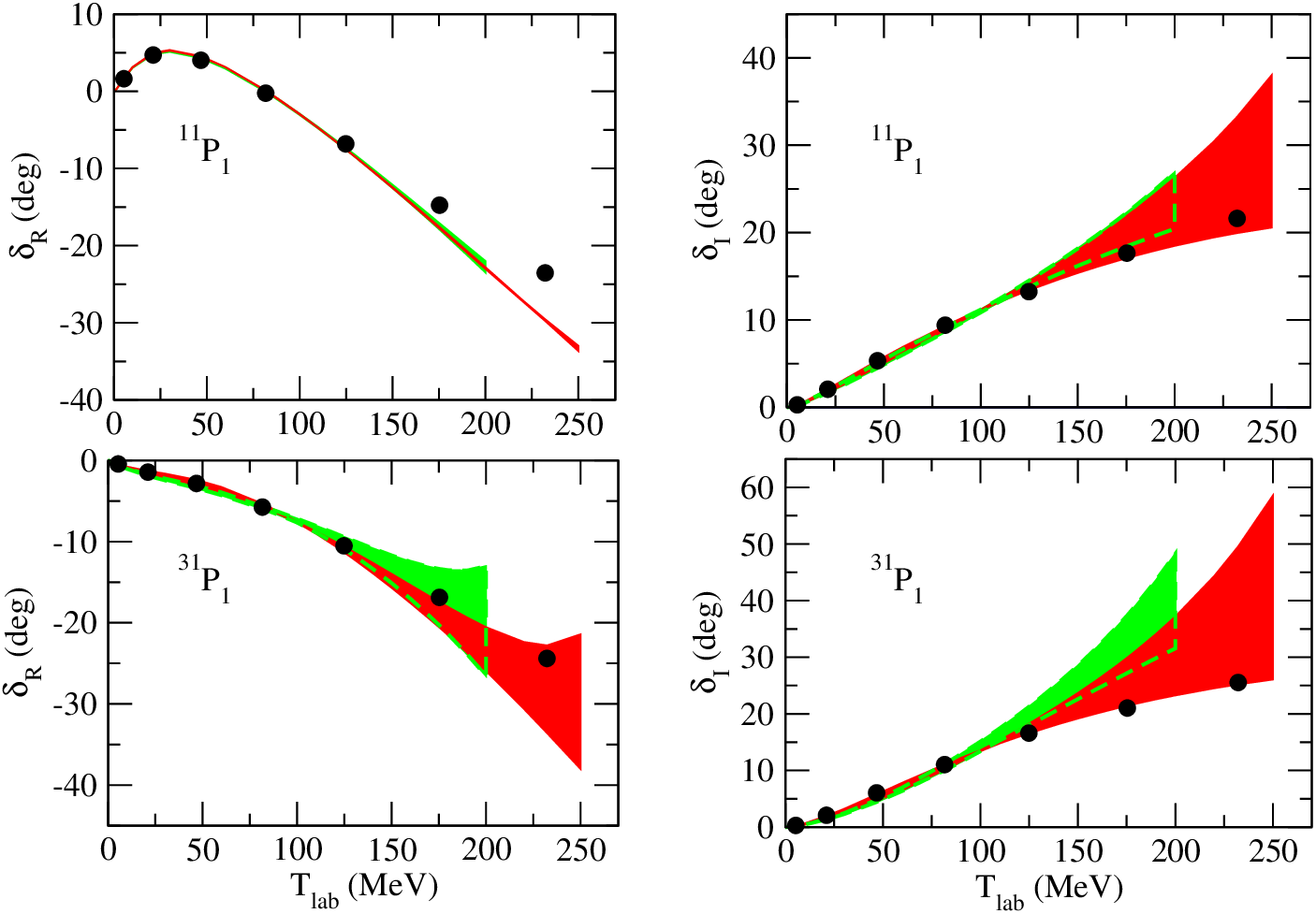}
\vskip 1.5cm
\includegraphics[height=85mm]{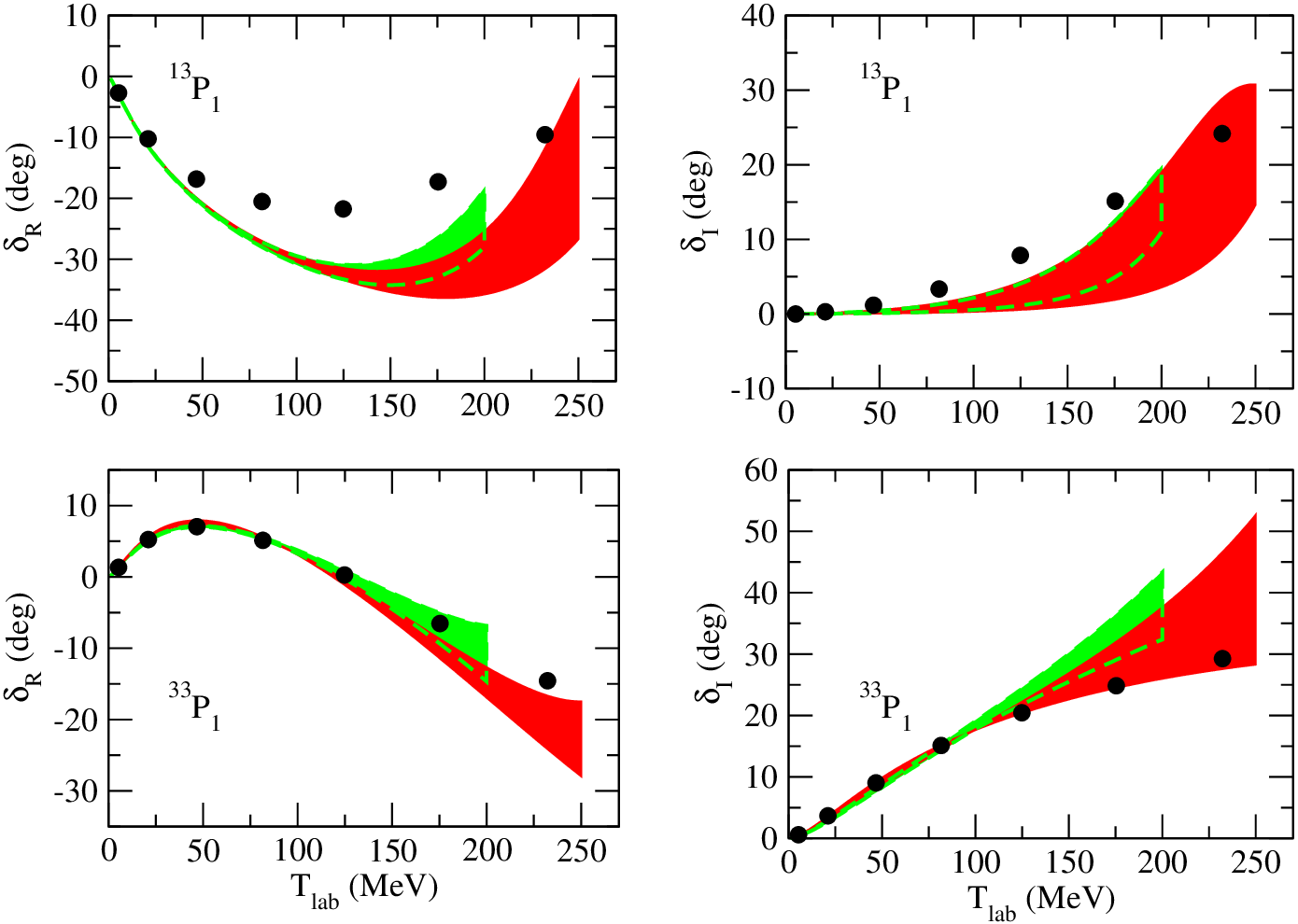}

\caption{Real and imaginary parts of the phase shift in the $^1P_1$ and $^3P_1$ partial waves.
The red/dark band shows the chiral EFT results up to NNLO for variations of the cutoff
in the range $\Lambda =$ 450--650~MeV in the Lippmann-Schwinger equation,
while the green/light band are results to NLO for $\Lambda =$ 450--600~MeV. The cutoff in the
pion loops is varied independently in the range $\tilde \Lambda =$ 500--700~MeV.
The solid circles represent the solution of the PWA of Ref.~\cite{Zhou2012}. 
}
\label{fig:1P1}
\end{center}
\end{figure}

\begin{figure}[h]
\begin{center}
\includegraphics[height=85mm]{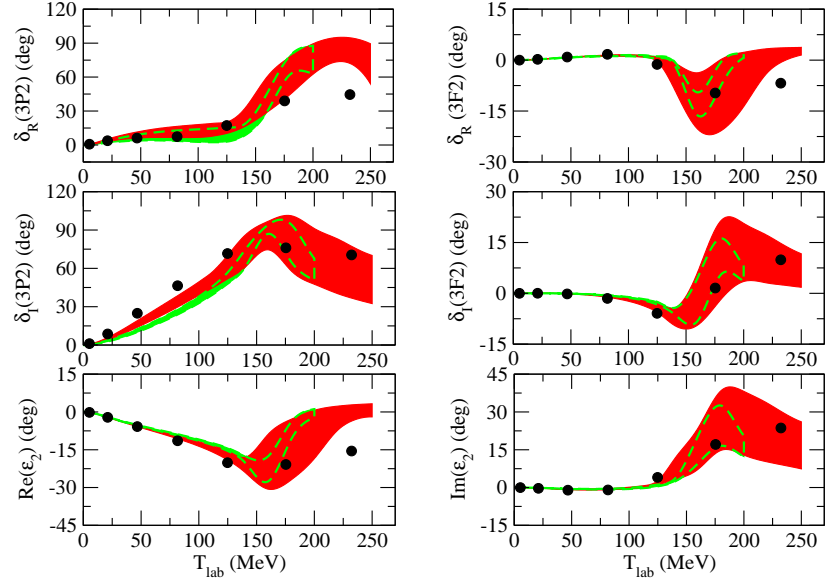}
\vskip 1.0cm
\includegraphics[height=85mm]{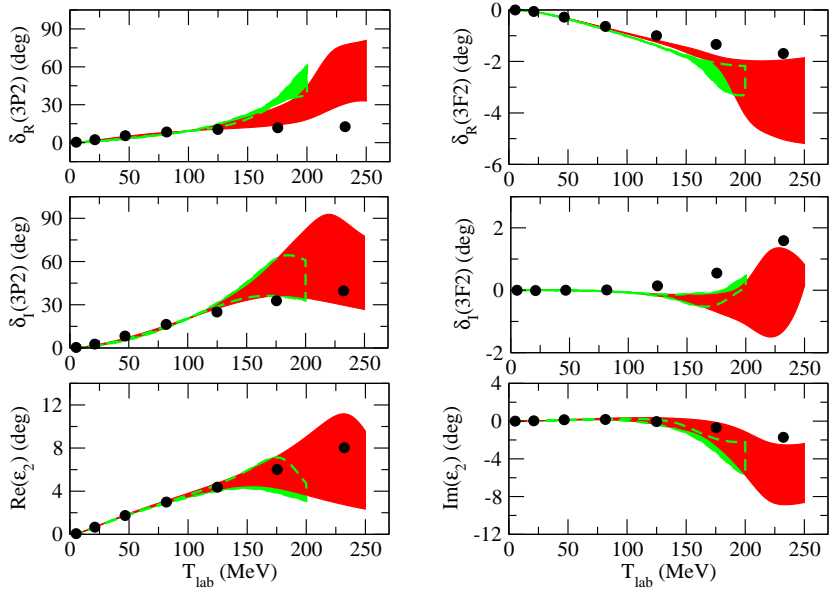}

\caption{Real and imaginary parts of the phase shift in the $^3P_2$--$^3F_2$ partial wave.
The red/dark band shows the chiral EFT results up to NNLO for variations of the cutoff
in the range $\Lambda =$ 450--650~MeV in the Lippmann-Schwinger equation,
while the green/light band are results to NLO for $\Lambda =$ 450--600~MeV. The cutoff in the
pion loops is varied independently in the range $\tilde \Lambda =$ 500--700~MeV.
The solid circles represent the solution of the PWA of Ref.~\cite{Zhou2012}.
}
\label{fig:3PF2}
\end{center}
\end{figure}

\begin{figure}[h]
\begin{center}
\includegraphics[height=85mm]{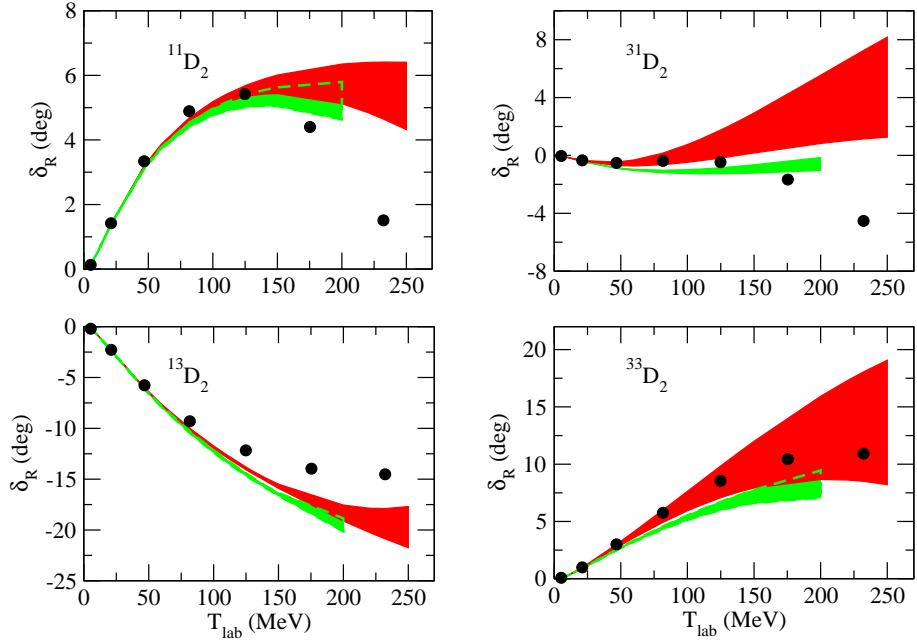}

\caption{Real part of the phase shift in the $^1D_2$ and $^3D_2$ partial waves. 
The red/dark band shows the chiral EFT results up to NNLO for variations of the cutoff
in the range $\Lambda =$ 450--650~MeV in the Lippmann-Schwinger equation,
while the green/light band are results to NLO for $\Lambda =$ 450--600~MeV. The cutoff in the
pion loops is varied independently in the range $\tilde \Lambda =$ 500--700~MeV.
The solid circles represent the solution of the PWA of Ref.~\cite{Zhou2012}.
}
\label{fig:D2}
\end{center}
\end{figure}

Our results are displayed and compared with the $\bar NN$ PWA \cite{Zhou2012}
in Figs.~\ref{fig:1S0}-\ref{fig:D2}. The bands represent the variation of the 
obtained phase shifts and mixing parameters with the cutoff.
Those variations can be viewed as an estimate for the theoretical uncertainty. 
Thus, in principle for the same variation of the cutoff those bands should
become narrower and narrower when one goes to higher order. However, as
argued in Ref.~\cite{Epe04a}, in practice one has to be careful in the
interpretation of the bands, specifically for the transition from NLO to NNLO.
Since the same number of contact terms are present in the interactions
at NLO and NNLO one rather should expect variations of similar magnitude.
In particular, for reasons discussed in \cite{Epe04a} the cutoff 
variation underestimates the uncertainty for the NLO results. In any
case one has to keep in mind that, following Ref.~\cite{Epe04a}, we
use a larger cutoff region at NNLO than for the NLO case. 

Let us now discuss the individual partial waves. 
Results for the $^1S_0$ channel
can be found in the upper part of Fig.~\ref{fig:1S0}. Obviously, the phase shift
for isospin $I=0$ (we use here the spectral notation $^{(2I+1)(2S+1)}L_J$)
is very well described up to fairly high energies -- even at NLO -- and likewise 
the inelasticity, presented in terms of the imaginary part of the phase shift. 
Moreover, the dependence on the cutoff is very small. 
In the $I=1$ channel the situation is rather different. Here we observe a
sizeable cutoff dependence of the results for energy above 150 MeV. This has 
to do with the fact that the PWA suggests a resonance-like behavior of the 
phase in this region. Since this resonance lies in an energy region where 
we expect our results to show increasing uncertainties, based on the experience
from the $NN$ case \cite{Epe04a}, it is not surprising that it is difficult 
to reproduce this structure quantitatively.
Nevertheless, there is a visible improvement when going from NLO to NNLO
and at the latter order the empirical phase shifts already lie within the error 
bands of theory. 

\begin{table}[H]
\begin{center}
\begin{tabular}{||cc||r|r|r|r||}
\hline \hline
& &  &  &  &  \\[-2.5ex]
 \multicolumn{2}{||c||}{LEC}   &  $\{ 450, \; 500\}$   &  $\{ 600, \; 500\}$  
&  $\{ 450, \; 700\}$  &  $\{ 600, \; 700\}$ \\[0.5ex]
\hline \hline
& &  &  &  &  \\[-2.5ex]
\multirow{6}{*}{$I=0$} & $\tilde C_{^1S_0}$ &  $-0.151$  & $-0.267$  &  $-0.151$  &  $-0.273$ \\[0.05ex]
& $C_{^1S_0}$        &   $0.455$  &  $0.436$  &   $0.454$  &   $0.426$ \\[0.05ex]
& $\tilde C^a_{^1S_0}$ &  $0.270$  & $0.232$  &  $0.232$  &  $0.177$ \\[0.05ex]
& $C^a_{^1S_0}$        &   $-0.915$  &  $-0.277$  &   $-0.905$  &   $-0.206$ \\[0.05ex]
& $C_{^3P_0}$ &  $1.150$  & $1.453$  &  $1.398$  &  $1.724$ \\[0.05ex]
& $C^a_{^3P_0}$ &  $0.769$  & $0.478$  &  $0.754$  &  $0.455$ \\[0.05ex]
& &  &  &  &  \\[-2.5ex]
\hline
& &  &  &  &  \\[-2.5ex]
\multirow{6}{*}{$I=1$} & $\tilde C_{^1S_0}$ &  $0$  & $0$  &  $0$  &  $0$ \\[0.05ex]
& $C_{^1S_0}$        &   $0.446$  &  $0.692$  &   $0.449$  &   $0.675$ \\[0.05ex]
& $\tilde C^a_{^1S_0}$ &  $1.329$  & $2.108$  &  $1.460$  &  $2.202$ \\[0.05ex]
& $C^a_{^1S_0}$        &   $-1.118$  &  $-0.369$  &   $-1.214$  &   $-0.498$ \\[0.05ex]
& $C_{^3P_0}$ &  $-0.357$  & $-0.074$  &  $-0.321$  &  $0.041$ \\[0.05ex]
& $C^a_{^3P_0}$ &  $0.501$  & $0.232$  &  $0.498$  &  $0.222$ \\[0.05ex]
&   &  &  &  &  \\[-2.5ex]
\hline\hline
&   &  &  &  &  \\[-2.5ex]
\multirow{10}{*}{$I=0$} & $C_{^1P_1}$ &  $0.384$  &   $-0.015$  &  $0.394$  &   $0.020$ \\[0.05ex]
& $C^a_{^1P_1}$ &  $0.711$  &   $0.714$  &  $0.709$  &   $0.705$ \\[0.05ex]
& $C_{^3P_1}$ &  $-0.374$  &   $-0.235$  &  $-0.296$  &   $-0.146$ \\[0.05ex]
& $C^a_{^3P_1}$ &  $0.381$  &   $0.190$  &  $0.378$  &   $0.194$ \\[0.05ex]
& $\tilde C_{^3S_1}$ &  $-0.132$  &   $-0.083$  &  $-0.122$  &   $-0.075$ \\[0.05ex]
& $C_{^3S_1}$        &   $-0.497$  &  $-0.623$  &   $-0.731$  &  $-0.853$ \\[0.05ex]
& $\tilde C^a_{^3S_1}$ &  $0.334$  &   $0.325$  &  $0.319$  &   $0.301$ \\[0.05ex]
& $C^a_{^3S_1}$        &   $0.221$  &  $-0.573$  &   $0.325$  &  $-0.438$ \\[0.05ex]
& $C_{\epsilon_{1}}$ &   $0.496$  &  $0.520$  &   $0.557$  &  $0.585$ \\[0.05ex]
& $C^a_{\epsilon_{1}}$ &   $-0.599$  &  $-0.218$  &   $-0.653$  &  $-0.290$ \\[0.05ex]
&   &  &  &  &  \\[-2.5ex]
\hline
& &  &  &  &  \\[-2.5ex]
\multirow{10}{*}{$I=1$} & $C_{^1P_1}$ &  $-0.623$  &   $-0.735$  &  $-0.659$  &   $-0.858$ \\[0.05ex]
& $C^a_{^1P_1}$ &  $0.682$  &   $0.544$  &  $0.688$  &   $0.573$ \\[0.05ex]
& $C_{^3P_1}$ &  $-0.180$  &   $-0.373$  &  $-0.201$  &   $-0.443$ \\[0.05ex]
& $C^a_{^3P_1}$ &  $0.716$  &   $0.628$  &  $0.719$  &   $0.645$ \\[0.05ex]
& $\tilde C_{^3S_1}$ &  $-0.089$  &   $-0.120$  &  $-0.087$  &   $-0.122$ \\[0.05ex]
& $C_{^3S_1}$        &   $0.698$  &  $0.148$  &   $0.707$  &  $0.188$ \\[0.05ex]
& $\tilde C^a_{^3S_1}$ &  $0.399$  &   $0.210$  &  $0.398$  &   $0.224$ \\[0.05ex]
& $C^a_{^3S_1}$        &   $0.164$  &  $0.665$  &   $0.124$  &  $0.602$ \\[0.05ex]
& $C_{\epsilon_{1}}$ &   $0.245$  &  $0.182$  &   $0.279$  &  $0.237$ \\[0.05ex]
& $C^a_{\epsilon_{1}}$ &   $0.015$  &  $0.111$  &   $-0.019$  &  $-0.046$ \\[0.05ex]
&   &  &  &  &  \\[-2.5ex]
\hline\hline
&   &  &  &  &  \\[-2.5ex]
\multirow{2}{*}{$I=0$} & $C_{^3P_2}$ &  $0.225$  &   $0.466$  &  $0.363$  &   $0.630$ \\[0.05ex]
& $C^a_{^3P_2}$ &  $0.674$  &   $0.428$  &  $0.661$  &   $0.410$ \\[0.05ex]
& &  &  &  &  \\[-2.5ex]
\hline
& &  &  &  &  \\[-2.5ex]
\multirow{2}{*}{$I=1$} & $C_{^3P_2}$ &  $-0.362$  &   $-0.268$  &  $-0.361$  &   $-0.266$ \\[0.05ex]
& $C^a_{^3P_2}$ &  $0.528$  &   $0.350$  &  $0.529$  &   $0.351$\\[0.5ex] 
\hline \hline
\end{tabular}
\vskip 0.2 true cm
\parbox{16.5cm}{
\caption{The LECs at NLO for the different cutoff combinations
$\big\{ \Lambda \, [\mbox{MeV}], \; \tilde \Lambda \, [\mbox{MeV}] \big\}$. 
The values of the $\tilde C_i$ are in unit of $10^4$ GeV$^{-2}$ and the $C_i$ in $10^4$ GeV$^{-4}$. 
The parameters related to annihilation, $\tilde C^a_i$ and $C^a_i$ 
(see Eqs.~(\ref{ANN1})--(\ref{ANN2})), are 
in units of $10^2$ GeV$^{-1}$ and $10^2$ GeV$^{-3}$, respectively. 
\label{tab:LEC1}}
}
\end{center}
\end{table}


\begin{table}[H]
\begin{center}
\begin{tabular}{||cc||r|r|r|r||}
\hline \hline
& &  &  &  &  \\[-2.5ex]
 \multicolumn{2}{||c||}{LEC}   &  $\{ 450, \; 500\}$   &  $\{ 650, \; 500\}$  
&  $\{ 450, \; 700\}$  &  $\{ 650, \; 700\}$ \\[0.5ex]
\hline \hline
& &  &  &  &  \\[-2.5ex]
\multirow{6}{*}{$I=0$} & $\tilde C_{ ^1S_0}$ &  $-0.140$  & $-0.278$  &  $-0.141$  &  $-0.299$ \\[0.05ex]
& $C_{^1S_0}$        &   $0.456$  &  $0.459$  &   $0.456$  &   $0.463$ \\[0.05ex]
& $\tilde C^a_{ ^1S_0}$ &  $0.208$  & $0.247$  &  $0.155$  &  $0.219$ \\[0.05ex]
& $C^a_{ ^1S_0}$        &   $-1.063$  &  $-0.337$  &   $-1.045$  &   $-0.233$ \\[0.05ex]
& $C_{ ^3P_0}$ &  $0.031$  & $0.310$  &  $-0.444$  &  $-0.217$ \\[0.05ex]
& $C^a_{ ^3P_0}$ &  $0.796$  & $0.492$  &  $0.828$  &  $0.556$ \\[0.05ex]
& &  &  &  &  \\[-2.5ex]
\hline
& &  &  &  &  \\[-2.5ex]
\multirow{6}{*}{$I=1$} & $\tilde C_{^1S_0}$ &  $0.025$  & $0.095$  &  $0.052$  &  $-0.011$ \\[0.05ex]
& $C_{^1S_0}$        &   $0.453$  &  $0.213$  &   $0.450$  &   $0.189$ \\[0.05ex]
& $\tilde C^a_{^1S_0}$ &  $1.884$  & $2.483$  &  $2.129$  &  $3.847$ \\[0.05ex]
& $C^a_{^1S_0}$        &   $-1.733$  &  $-2.778$  &   $-2.566$  &   $-4.474$ \\[0.05ex]
& $C_{^3P_0}$ &  $-0.535$  & $-0.117$  &  $-0.531$  &  $-0.116$ \\[0.05ex]
& $C^a_{^3P_0}$ &  $0.514$  & $0.182$  &  $0.517$  &  $0.182$ \\[0.05ex]
&   &  &  &  &  \\[-2.5ex]
\hline\hline
&   &  &  &  &  \\[-2.5ex]
\multirow{10}{*}{$I=0$} & $C_{^1P_1}$ &  $0.400$  &   $-0.113$  &  $0.438$  &   $-0.069$ \\[0.05ex]
& $C^a_{^1P_1}$ &  $0.722$  &   $0.637$  &  $0.721$  &   $0.634$ \\[0.05ex]
& $C_{^3P_1}$ &  $-0.521$  &   $-0.339$  &  $-0.596$  &   $-0.432$ \\[0.05ex]
& $C^a_{^3P_1}$ &  $0.417$  &   $0.168$  &  $0.421$  &   $0.175$ \\[0.05ex]
& $\tilde C_{^3S_1}$ &  $-0.162$  &   $-0.100$  &  $-0.183$  &   $-0.103$ \\[0.05ex]
& $C_{^3S_1}$        &   $0.353$  &  $0.204$  &   $0.728$  &  $0.526$ \\[0.05ex]
& $\tilde C^a_{^3S_1}$ &  $0.364$  &   $0.371$  &  $0.397$  &   $0.415$ \\[0.05ex]
& $C^a_{^3S_1}$        &   $0.087$  &  $-0.841$  &   $-0.117$  &  $-1.125$ \\[0.05ex]
& $C_{\epsilon_{1}}$ &   $0.205$  &  $0.236$  &   $0.062$  &  $0.106$ \\[0.05ex]
& $C^a_{\epsilon_{1}}$ &   $-0.485$  &  $-0.002$  &   $-0.362$  &  $0.167$ \\[0.05ex]
&   &  &  &  &  \\[-2.5ex]
\hline
& &  &  &  &  \\[-2.5ex]
\multirow{10}{*}{$I=1$} & $C_{^1P_1}$ &  $-1.013$  &   $-1.294$  &  $-1.349$  &   $-1.869$ \\[0.05ex]
& $C^a_{^1P_1}$ &  $0.711$  &   $0.535$  &  $0.775$  &   $0.668$ \\[0.05ex]
& $C_{^3P_1}$ &  $-0.530$  &   $-0.902$  &  $-0.794$  &   $-1.356$ \\[0.05ex]
& $C^a_{^3P_1}$ &  $0.742$  &   $0.630$  &  $0.788$  &   $0.735$ \\[0.05ex]
& $\tilde C_{^3S_1}$ &  $-0.067$  &   $-0.143$  &  $-0.044$  &   $-0.125$ \\[0.05ex]
& $C_{^3S_1}$        &   $1.150$  &  $0.764$  &   $1.325$  &  $1.235$ \\[0.05ex]
& $\tilde C^a_{^3S_1}$ &  $0.413$  &   $0.282$  &  $0.411$  &   $0.402$ \\[0.05ex]
& $C^a_{^3S_1}$        &   $-0.336$  &  $0.211$  &   $-0.896$  &  $-0.441$ \\[0.05ex]
& $C_{\epsilon_{1}}$ &   $0.320$  &  $0.287$  &   $0.376$  &  $0.383$ \\[0.05ex]
& $C^a_{\epsilon_{1}}$ &   $-0.065$  &  $0.021$  &   $-0.182$  &  $-0.162$ \\[0.05ex]
&   &  &  &  &  \\[-2.5ex]
\hline\hline
&   &  &  &  &  \\[-2.5ex]
\multirow{2}{*}{$I=0$} & $C_{^3P_2}$ &  $-0.300$  &   $-0.120$  &  $-0.518$  &   $-0.399$ \\[0.05ex]
& $C^a_{^3P_2}$ &  $0.707$  &   $0.402$  &  $0.731$  &   $0.443$ \\[0.05ex]
& &  &  &  &  \\[-2.5ex]
\hline
& &  &  &  &  \\[-2.5ex]
\multirow{2}{*}{$I=1$} & $C_{^3P_2}$ &  $-0.648$  &   $-0.558$  &  $-0.821$  &   $-0.782$ \\[0.05ex]
& $C^a_{^3P_2}$ &  $0.544$  &   $0.329$  &  $0.565$  &   $0.377$\\[0.5ex] 
\hline \hline
\end{tabular}
\vskip 0.2 true cm
\parbox{16.5cm}{
\caption{The LECs at NNLO 
for the different cutoff combinations
$\big\{ \Lambda \, [\mbox{MeV}], \; \tilde \Lambda \, [\mbox{MeV}] \big\}$. 
The values of the $\tilde C_i$ are in unit of $10^4$ GeV$^{-2}$ and the $C_i$ in $10^4$ GeV$^{-4}$.
The parameters related to annihilation, $\tilde C^a_i$ and $C^a_i$
(see Eqs.~(\ref{ANN1})--(\ref{ANN2})), are 
in units of $10^2$ GeV$^{-1}$ and $10^2$ GeV$^{-3}$, respectively. 
\label{tab:LEC2}}
}
\end{center}
\end{table}

\begin{figure}[h]
\begin{center}
\includegraphics[height=85mm]{13SD1.eps}
\vskip 1.0cm
\includegraphics[height=85mm]{33SD1.eps}

\caption{Real and imaginary parts of the phase shift in the $^3S_1$--$^3D_1$ partial wave.
The red/dark band shows the chiral EFT results up to NNLO for variations of the cutoff
in the range $\Lambda =$ 450--650~MeV in the Lippmann-Schwinger equation,
while the green/light band are results to NLO for $\Lambda =$ 450--600~MeV. The cutoff in the
pion loops is varied independently in the range $\tilde \Lambda =$ 500--700~MeV.
The solid circles represent the solution of the PWA of Ref.~\cite{Zhou2012}. 
}
\label{fig:3SD1}
\end{center}
\end{figure}

We want to emphasize that this improvement is entirely due to 
inclusion of the subleading two-pion exchange potential, since as already stressed
above no new contact terms arise at NNLO and thus the number of adjustable 
parameters is the same at 
NLO and NNLO. Also, it should be said that the NLO result, shown here up to
$T_{lab}=200$ MeV, exhibits a similar trend like the one for NNLO  
at higher energies, i.e. the phases reach a maximum and then become more
negative again.

The situation for the $^3P_0$ partial wave is similar, see. Fig.~\ref{fig:1S0}
(lower part). 
Also here the $I=0$ phase shifts are well reproduced while in the $I=1$
case there is an even larger cutoff dependence than in the $^{31}S_0$. 
Obviously also the $^{33}P_0$ amplitude of the PWA \cite{Zhou2012} exhibits 
a resonance-like behavior. 
Its reproduction requires a potential that is repulsive at large separations 
of the antinucleon and nucleon but becomes attractive for short distances. 
Since there is only a single LEC up to NNLO for $P$ waves, the magnitude and 
range of such an attraction cannot be adequately accounted for. 
For improvements one has to wait for a N$^3$LO calculation.

Results for the $^1P_1$ and $^3P_1$ partial waves are shown in Fig.~\ref{fig:1P1}.
In general, the description improves when going from NLO to NNLO. Specifically 
for the two $^1P_1$ channels and the $^{33}P_1$ the results at NNLO agree with 
those of the PWA within the uncertainty bands for energies up to 150 MeV and
often even up to 250 MeV. 
An exception is the $^{13}P_1$ partial wave where the phase shift 
can only be described up to 50 MeV or so. 
Similar to the $^{33}P_0$, the PWA yields a negative phase at low energies 
which tends towards positive values at larger energies \cite{Zhou2012} and 
one encouters the same difficulty as discussed above. 

In Fig.~\ref{fig:3SD1} one can find our results for the coupled 
$^3S_1$--$^3D_1$ partial wave. Here the $S$-wave phase shifts (and also
the inelasticity) are satisfactorily described over the whole energy
range considered with uncertainties comparable to those observed for
the $NN$ interaction \cite{Epe04a}. There is a larger cutoff dependence in 
the $D$ waves and the mixing parameter $\epsilon_1$, specifically for $I=0$. 
However, one has to keep in mind that there is no LEC up to NNLO for the $D$ 
waves. The $^{33}D_1$ exhibits the trend of turning from negative to positive 
values at higher energies which cannot be described in an NNLO calculation,
as discussed above. 

The situation in the $^3P_2$--$^3F_2$ channel is displayed in Fig.~\ref{fig:3PF2}.
In general our results agree with those of the PWA up to about 200 MeV within
the uncertainty. Stronger deviations are visible again for those phases
which show a resonance-like behavior like, e.g., the $^{13}P_2$.

At last, in Fig.~\ref{fig:D2} the $^1D_2$ and $^3D_2$ phase shifts are 
presented. There are no LECs in those partial waves up to NNLO and, 
thus, our results are genuine predictions. The potential consists only of 
one- and two-pion exchange and, consequently, there is no contribution to 
annihilation. Thus, $\delta_I \equiv 0$ and we do not show this quantity.

\begin{table}[htb]
\vspace{0.6cm}
\begin{center}
\begin{tabular}{||cl||c|c||}
\hline \hline
  \multicolumn{2}{||c||}{ } &{}   &{}\\[-1.5ex]
  \multicolumn{2}{||c||}{ } & I=0  &I=1\\[1ex]
\hline\hline
  \multicolumn{2}{||c||}{ } &{}   &{}\\[-1.5ex]
  \multirow{3}{*}{$^1S_0$}  &NLO &$-$0.21\,$-$\,i\,(1.20 $\ccc$ 1.21) 
&(1.03 $\ccc$ 1.04)\,$-$\,i\,(0.56 $\ccc$ 0.58)\\[1ex]
  &NNLO &$-$0.21\,$-$\,i\,(1.21 $\ccc$ 1.22)  &(1.02 $\ccc$  1.04)\,$-$\,i\,(0.57 $\ccc$ 0.61)\\[1ex]
  &model D &$-$0.23\,$-$\,i\,1.01  &0.99\,$-$\,i\, 0.58\\
\hline
  \multicolumn{2}{||c||}{ } &{}   &{}\\[-1.5ex]
 \multirow{3}{*}{$^3S_1$}  &NLO &(1.34 $\ccc$ 1.37)\,$-$\,i\,(0.88 $\ccc$ 0.90) &(0.43 $\ccc$ 0.44)\,$-$\,i\,(0.87 $\ccc$ 0.90) \\[1ex]
   &NNLO  &(1.37 $\ccc$ 1.38)\,$-$\,i\,(0.86 $\ccc$ 0.88)  &(0.43 $\ccc$ 0.44)\,$-$\,i\,(0.91 $\ccc$ 0.92)\\[1ex]
   &model D &1.55\,$-$\,i\,1.45  &0.33\,$-$\,i\,0.96\\[1ex]
\hline  \hline
  \end{tabular}
\vspace{0.3cm}
\caption{Scattering lengths (in fm) for the $^1S_0$ and $^3S_1$ partial waves 
in the isospin $I=0$ and $I=1$ channels. Results based on the NLO and NNLO 
potentials are given and compared with the predictions of the 
J\"ulich $\bar NN$ model D \cite{Mull}.
}\label{tab2}
\vspace{-5mm}
\end{center}
\end{table}

Results for the scattering lengths in the $^1S_0$ and $^3S_1$ partial 
waves are summarized in Table~\ref{tab2}. These are complex numbers
because of the presence of annihilation. The scattering lengths 
implied directly by the PWA of \cite{Zhou2012} are not provided in 
that reference. Thus, the lowest energy that enters our fitting procedure
concerns the phase shifts at $p_{lab} = 100$~MeV/c which corresponds 
to $T_{lab} = 5.3$~MeV. In view of that one can consider our values 
as predictions of chiral EFT. 
As one can see in Table~\ref{tab2} we get practically the same results at
NLO and at NNLO and, moreover, there is very little cutoff dependence. 
Actually, in case of $\textrm{Re}\,a_{^1S_0}$ in the $I=0$ channel there is 
no variation in the first two digits and, therefore, only a single number 
is given.

Table~\ref{tab2} contains also scattering lengths predicted by the 
most refined meson-exchange potential developed by the J\"ulich group,
namely model D published in \cite{Mull}. It is interesting to see that
the results are very similar not only on a qualitative level but in most 
cases even on a quantitative level. One has to keep in mind that there 
are no data that would allow one to fix the relative magnitude of the 
singlet- and triplet- contributions near threshold. 
Moreover, the
J\"ulich $\bar NN$ potential was only fitted to integrated cross sections.
Differential cross sections or polarization data were not considered. 

There is some experimental information that puts constraints on these 
scattering lengths. 
Measurements of the level shifts and widths of antiproton-proton allow
one to deduce values for the spin-averaged $\bar pp$ scattering lengths
via the Deser-Trueman formula. 
Corresponding results taken from Ref.~\cite{Gotta} are listed in Table~\ref{tab3}. 
In that reference one can also find values for the imaginary part of the 
scattering lengths that are deduced from measurements of the ($\bar np$ 
and $\bar pp$) annihilation cross section.

\begin{table}[htb]
\vspace{0.6cm}
\begin{center}
\begin{tabular}{||c||lc|c|c||}
\hline \hline 
 {} &\multicolumn{2}{c|}{ } & {}  &{}\\[-1.5ex]
 {} &\multicolumn{2}{|c|}{chiral EFT} & model D & Experiment\\[1ex]
\hline\hline
 {} &\multicolumn{2}{c|}{ } & {}  &{}\\[-1.5ex]
   \multirow{4}{*}{$\bar a_{S,\bar pp}$} &NLO &\phantom{$+$\,1\,}(0.77 $\ccc$ 0.79) 
            &\multirow{4}{*}{0.80\,$-$\,i\,1.10} &{} \\
   {} & {} &$-$\,i\,(0.88 $\ccc$ 0.90) &{} &\phantom{$+$\,1\,}{(0.95 $\pm$ 0.02)}\\[1ex]
   {} & NNLO &\phantom{$+$\,1\,}(0.78 $\ccc$ 0.79) &{} &{$-$\,i\,(0.73 $\pm$ 0.03)} \\
   {} &{} &$-$\,i\,(0.89 $\ccc$ 0.91) &{} &{}\\[1ex]
\hline
   {} &\multicolumn{2}{c|}{ } & {}  &{}\\[-1.5ex]
  \multirow{2}{*}{Im $\bar a_{S,I=1}$}  &NLO &($-$0.82 $\ccc$ $-$0.79)  &\multirow{2}{*}{$-$0.86} 
                                           &\multirow{2}{*}{($-$0.83 $\pm$ 0.07)} \\[1ex]
   {} &NNLO &($-$0.84 $\ccc$ $-$0.83)  &{}   & {}\\[1ex]
\hline

  {} &\multicolumn{2}{c|}{ } & {}  &{}\\[-1.5ex]
  \multirow{2}{*}{Im $\bar a_{S,I=0}$}  &NLO &($-$0.98 $\ccc$ $-$0.96) &\multirow{2}{*}{$-$1.34}
                                           &\multirow{2}{*}{($-$0.63 $\pm$ 0.08)} \\[1ex]
   {} &NNLO &($-$0.97$\ccc$ $-$0.95)  &{}   & {}\\[1ex]
\hline  \hline
  \end{tabular}
\vspace{0.3cm}
\caption{Spin-averged $S-$wave scattering length (in fm). 
Results based on the NLO and NNLO potentials are given and compared with the 
predictions of the J\"ulich $\bar NN$ model D \cite{Mull}.
The experimental information is taken from Ref.~\cite{Gotta}. 
}\label{tab3}
\vspace{-5mm}
\end{center}
\end{table} 

As far as we know, this experimental evidence was not taken into
account in the PWA \cite{Zhou2012}. Nonetheless, for completeness
we provide the predictions based on our EFT interaction. One should be
cautious, however, in comparing our results with the experimental numbers.
As said above, our calculations are performed in the isospin basis
so that $a_{\bar pp}$ is simply given by $(a_{I=0}+a_{I=1})/2$. 
It is known that the presence of the Coulomb force in $\bar pp$ and 
the $p$-$n$ mass difference lead to changes of the $S$-wave 
scattering lengths in the order of $0.1$~fm \cite{Carbonell} 
and, therefore, one should not take quantitative differences too serious.

Finally, let us discuss $\bar NN$ bound states. Several of the phase shifts
tabulated in Ref.~\cite{Zhou2012} start at 180$^\circ$ at \mbox{$T_{lab} = 0$~MeV},
namely $^{11}S_0$, $^{13}P_0$, $^{13}S_1$, and $^{33}S_1$, 
which according to the standard convention based on the Levinson theorem 
signals the presence of a bound state. Therefore, we performed
a search for possible bound states generated by our EFT interaction 
where we restricted ourselves to energies not too far from the $\bar NN$
threshold. We did not find any near-threshold poles in the
$^{11}S_0$ and $^{33}S_1$--$^{33}D_1$ partial waves. In case of the 
$^{13}S_1$--$^{13}D_1$ interaction there is a pole which corresponds to
a ``binding'' energy of $E_B = +(5.6\ccc 7.7)-{\rm i}\,(49.2\ccc 60.5)$ MeV, 
depending on the cutoffs $\{\Lambda,\tilde\Lambda\}$, at NLO and 
$E_B= +(4.8\ccc 21.3)-{\rm i}\,(60.6\ccc 74.9)$ MeV at NNLO. 
The positive sign of the real part of $E_B$ indicates that the poles
we found are actually located above the $\bar NN$ threshold. But they
move below the threshold when we switch off the imaginary part of the
potential and that is the reason why we refer to them as bound states. 
To be precise these are unstable bound states in the terminology of 
Ref.~\cite{Badalyan82}. Note that those poles lie on the physical sheet and, 
therefore, do not correspond to resonances. 
Evidently, the width of the state, $\Gamma = -2\, {\rm Im}\, E_B$, is
rather large.  
There is also a pole in the $^{13}P_0$ partial wave. It corresponds to
a binding energy of $E_B=(-1.1\ccc +1.9)-{\rm i}\,(17.8\ccc 22.4)$ MeV
at NLO and $E_B=-(3.7\ccc 0.2)-{\rm i}\,(22.0\ccc 26.4)$ MeV at NNLO.
In this context we want to mention that bound states and also resonances
have been likewise found in other studies of the $\bar NN$ interaction, 
see Refs.~\cite{Entem06,Bennich} for recent examples.

\section{Summary and outlook}
\label{sec:5}

In this paper we presented an exploratory study of the $\bar NN$ interaction 
in a chiral effective field theory approach based on a modified Weinberg power
counting, analoguous to the $NN$ case in \cite{Epe05,Epe04a}. 
The $\bar NN$ potential has been evaluated up to NNLO in the perturbative
expansion and the arising low-energy constants have been fixed by a
fit to the phase shifts and inelasticities provided by a recently
published phase-shift analysis of $\bar pp$ scattering data \cite{Zhou2012}. 
It turned out that the overall quality of the description of the 
$\bar NN$ amplitudes that can be achieved at NNLO is comparable to 
the one found in case of the $NN$ interaction at the same
order \cite{Epe04a}. 
Specifically, for the $S$-waves ($^{11}S_0$, $^{13}S_1$, $^{33}S_1$) 
nice agreement with the phase shifts and inelasticities of \cite{Zhou2012} 
has been obtained up to laboratory energies of about 200 MeV, i.e. over 
almost the
whole energy region considered. The same is also the case for many 
of the $P$-waves. Thus, we conclude that the chiral EFT approach,
applied successfully in Refs.~\cite{Entem:2003ft,Epe05} to the $NN$
interaction and in Refs.~\cite{Polinder06,JH13} to the hyperon-nucleon 
interaction, is very well suited for studies of the $\bar NN$ interaction too. 
 
Of course, there are also some visible deficiencies in our results.
They occur primarily in those partial waves where 
the partial-wave analysis of \cite{Zhou2012} suggests 
the presence of (presumably strongly inelastic) resonances at energies around
$T_{lab}\approx 200-250$~MeV. It is not surprising that structures in this
energy region cannot be reproduced reliably within our NNLO calculation. 
Clearly, here an extension of our investigation to N$^3$LO is necessary 
for improving the description of the $\bar NN$ interaction.
Therefore, we plan to extend our study to N$^3$LO in the future. At this
stage it will become sensible to perform the calculation in particle 
basis so that the Coulomb interaction in the $\bar pp$ system can 
be taken into account rigorously, and to compute observables and
compare them directly with scattering data for $\bar pp$
elastic scattering and for the charge-exchange reaction 
$\bar pp \to \bar nn$.
Annihilation processes that occur predominantly
at short distances reduce the magnitude of the $S$-wave amplitudes 
so that higher partial waves start to become import at much lower 
energies as compared to what one knows from the $NN$ interaction. 
Thus, without a realistic description of higher partial waves, and
particularly of the $D$-waves, it is not meaningful to confront 
the amplitudes resulting from our NNLO interaction directly with 
$\bar NN$ data and, therefore, we have refrained from doing so in 
the present work. 

\section*{Acknowledgements}

This work is supported in part by the DFG and the NSFC through
funds provided to the Sino-German CRC 110 ``Symmetries and
the Emergence of Structure in QCD'' and by the EU Integrated 
Infrastructure Initiative HadronPhysics3.

\end{document}